\documentstyle[12pt,aaspp4,psfig]{article} 

\begin{document}
\title{Discovery of the 198 s X-ray Pulsar GRO J2058+42}
\author
{Colleen~A.~Wilson, Mark~H.~Finger\altaffilmark{1}, B.~Alan~Harmon}
\affil{\footnotesize ES 84 Space Sciences Laboratory, NASA/Marshall
Space Flight Center, Huntsville, AL 35812;
wilsonc@gibson.msfc.nasa.gov, finger@gibson.msfc.nasa.gov,
harmon@ssl.msfc.nasa.gov}
\author{Deepto~Chakrabarty}
\affil{\footnotesize  Center for Space Research, Massachusetts
Institute of Technology, Cambridge, MA 02139; deepto@space.mit.edu} 
\altaffiltext{1}{{\em CGRO} Science Support Center/ Goddard Space Flight Center}
\and
\author{Tod Strohmayer}
\affil{\footnotesize Universities Space Research Association, Laboratory for
High Energy Astrophysics, NASA/Goddard Space Flight Center, Greenbelt, MD 
20771; stroh@pcasrv1.gsfc.nasa.gov}

\begin{abstract}

GRO J2058+42, a transient 198 second x-ray pulsar, was discovered by the Burst
and Transient Source Experiment (BATSE) on the {\em Compton Gamma-Ray 
Observatory (CGRO)}, during a ``giant" outburst in 1995 September-October. The 
total flux peaked at about 300 mCrab (20-50 keV) as measured by Earth 
occultation.  The pulse period decreased from 198 s to 196 s during the 46-day
outburst. The pulse shape evolved over the course of the outburst and exhibited
energy dependent variations. BATSE observed five additional weak 
outbursts from GRO J2058+42, each with two week duration and peak pulsed flux of
about 15 mCrab (20-50 keV), that were spaced by about 110 days. An observation
of the 1996 November outburst by the {\em Rossi X-ray Timing Explorer 
(RXTE)} Proportional Counter Array (PCA) localized the source to within 
a 4\arcmin\ radius error circle (90\% confidence) centered on R.A. = 20$^h$ 
59$^m$.0, Decl. = 41\arcdeg 43\arcmin\ (J2000). Additional shorter outbursts 
with peak pulsed fluxes of about 8 mCrab were detected by BATSE halfway between
the first four 15 mCrab outbursts.  The {\em RXTE} All-Sky Monitor detected all
8 weak outbursts with approximately equal durations and intensities. GRO J2058+42
is most likely a Be/X-ray binary that appears to outburst at periastron and
apastron. No optical counterpart has been identified 
to date and no x-ray source was present in the error circle in archival 
{\em ROSAT} observations.
\end{abstract}


\section{Introduction} 
In the last 25 years more than 40 accretion-powered x-ray pulsars have been
detected.  About half of these are transient, of which 12 have known Be star
companions.  Neutron stars with Be companions accrete material from the slow,
dense, stellar outflow thought to be confined to the equatorial plane of the Be
star.  Recent long term studies by the Burst and Transient Source Experiment 
(BATSE) on the {\em Compton Gamma-Ray Observatory (CGRO)} revealed that Be/X-ray
binaries exhibit series of often periodic outbursts.  These outbursts are
sometimes associated with ``giant'' outbursts accompanied by high
spin-up rates and luminosities.  BATSE observed four additional accreting
x-ray pulsars, without identified companions, which are believed to also be Be/X-ray binaries because their
temporal behavior closely resembles that of systems with Be companions
(\cite{Bildsten97}).  This paper reports the discovery and temporal behavior of
the fourth member of this group, GRO J2058+42.

A 198 second periodic signal was observed in the BATSE data starting
on 1995 September 14. At the same time a new source was also detected by 
Earth occultation measurements, which measure phase-averaged (total) flux. A 
location was determined from both the pulsed data and the non-pulsed data with
 a 95\% confidence error box of about 4\arcdeg\ $\times$ 1\arcdeg\ (\cite{Wilson95}). A {\em CGRO} 
target of opportunity was declared and the spacecraft was reoriented to allow
scans of the region by the Oriented Scintillation Spectroscopy Experiment
(OSSE), resulting in an improved 30\arcmin\ $\times$ 60\arcmin\ (95\%
confidence) position (\cite{Grove95}). The total flux peaked at about 300 mCrab
(20-100 keV) and the pulsed flux (RMS deviation from mean) peaked at 140 mCrab
(20-50 keV) on 1995 September 27. This bright outburst continued until 1995 October 30.  A search of archival 
BATSE data from 1991 April until this bright outburst showed no previous 
outbursts.

An analysis of BATSE data following the bright outburst initially revealed three much weaker
outbursts each lasting about two weeks with pulsed flux peaking at 15-20 mCrab 
(20-50 keV). These outbursts were spaced by about 110 days which allowed the 
peak of the next outburst to be predicted.  A target of opportunity scan of the
OSSE/BATSE error box was performed on 1996 November 28 with the {\em RXTE} 
Proportional Counter Array (PCA) yielding a 90\% confidence 4\arcmin\ radius error 
circle centered on R.A. = 20$^h$ 59$^m$.0, Decl. = +41\arcdeg 43\arcmin\
(\cite{Wilson96}). BATSE also detected the source from 1996 November 23-
December 1.  Another outburst was detected by BATSE about 110 days later (1997 
March 16-20.)  Three shorter outbursts with peak pulsed fluxes of about 8
mCrab were detected by BATSE halfway between the first four 15-20 mCrab 
outbursts.  The {\em RXTE} All-Sky Monitor detected all 8 weak outbursts with 
approximately equal durations and intensities. An archival search of 
{\em ROSAT} data found no sources within the error circle (J. Greiner, 1997, 
private communication.)  No optical counterpart has been found to date.

In this paper we present the BATSE and {\em RXTE} observations of GRO J2058+42.
Our observations with BATSE include histories of pulse frequency and pulsed 
flux from 1995 September - 1997 March,  a history of phase-averaged flux for the
``giant" outburst (1995 September-October), and pulse profile variations 
dependent upon energy and outburst phase of the ``giant" outburst. 
An {\em RXTE} PCA observation of a weak outburst on 1996 November 28 includes
a fit to the scan data used to better locate GRO J2058+42 and pulse profiles.
We compare BATSE and {\em RXTE} ASM observations of the 8 weak outbursts from
1995 December - 1997 March.  We then discuss the implications of our results. 

\section{Observations and Analyses}

BATSE consists of eight identical uncollimated detector modules positioned on 
the corners of the {\em CGRO} spacecraft such that the normal vectors of the 
detectors are perpendicular to the faces of a regular octahedron, providing
all-sky coverage.  The BATSE data presented here are taken with the large-area
detectors (LADs), which are NaI(Tl) scintillation crystals with a geometric 
area of 2025 cm$^2$ and a thickness of 1.27 cm.  The LADs are sensitive to 
photons from 20 to 1800 keV.  Two BATSE data types were used in this analysis, the CONT (2.048 s, 16 energy
channel) data and the DISCLA (1.024 s, 4 energy channel) data.  A more complete
description of the instrument and data types can be found in \cite{Fishman89}.

\subsection{The Initial ``Giant'' Outburst \label{sec:giant}}

\begin{figure}
\psfig{file=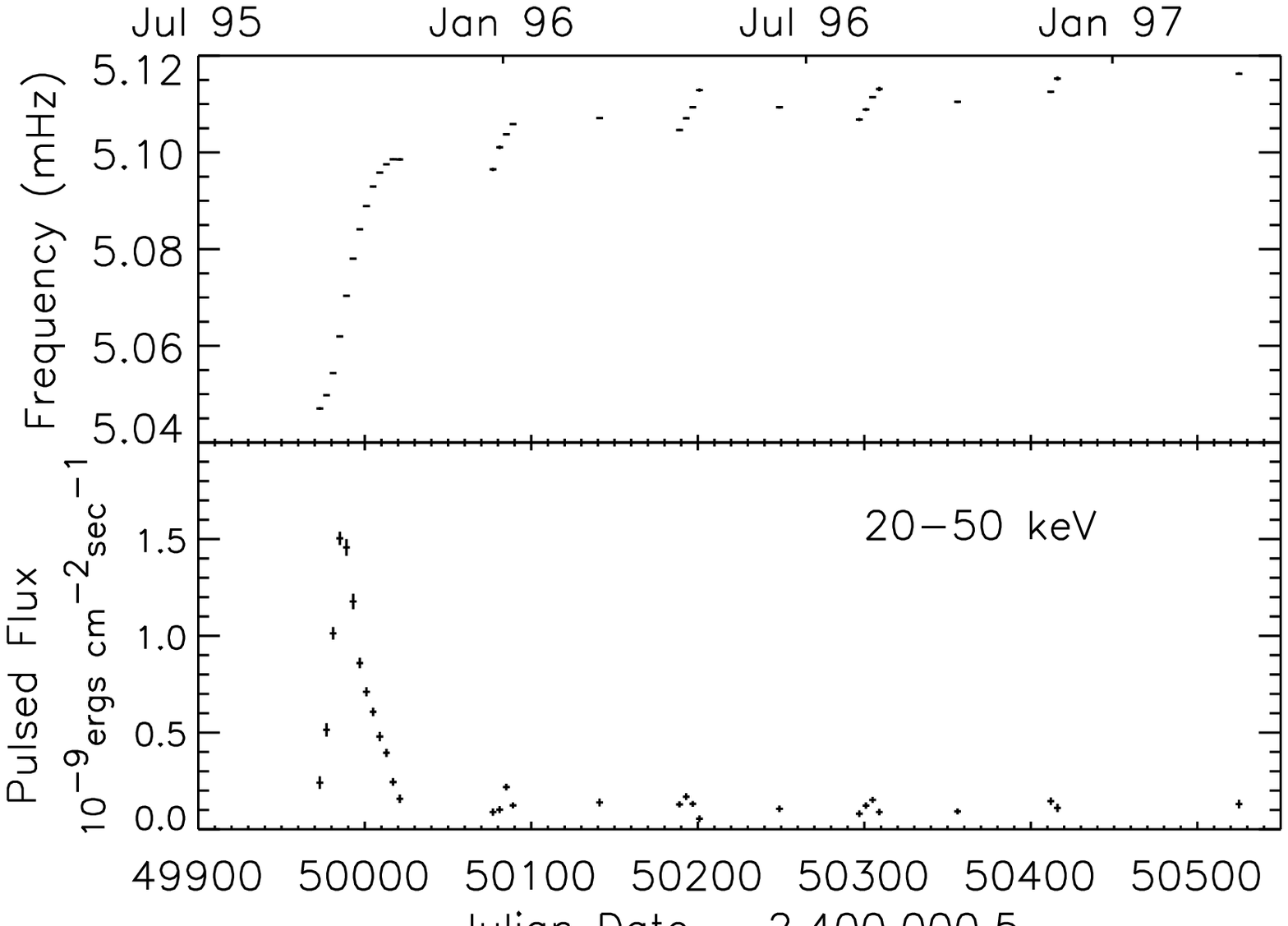,height=2.5in,width=4.0in}
\caption{GRO J2058+42 spin frequency and pulsed flux measurements from 
BATSE. The frequencies were determined at 4-day intervals from 20-50 keV
DISCLA data. The spin frequencies were barycentered, but have not been 
orbitally corrected since the orbit is unknown. Orbital effects are likely to
be a substantial part of the observed spin-up (or down) during and between the 
weak outbursts. The pulsed fluxes were 
determined at 4-day intervals by assuming an exponential spectrum
(see eq.~\protect\ref{eqn:spec}) with an e-folding energy of 20 keV. Upper limits
on the pulsed flux are not shown. \label{fig:freqhis}}
\end{figure}

Figure~\ref{fig:freqhis} shows histories of spin frequency and pulsed
intensity, determined by fits to 4-day intervals of the BATSE 20-50 keV DISCLA
(1.024 second) data.  Short intervals of data (200 s) were fit with a 
background model plus a sixth order Fourier expansion in the pulsed phase model
to generate pulse profiles at the model frequency. (See \cite{Bildsten97} for a
detailed description of pulsed flux and pulsed frequency estimation 
techniques.) The pulsed phase model initially consisted of a constant frequency
obtained from daily power spectra of the BATSE DISCLA data. The data were 
combined into 600 second intervals with continuous background models.  For each
four day span, the pulse profiles from the 600 s intervals were shifted in 
phase according to a range of trial frequency offsets from the pulse phase 
model ($\pm$ 5 cycles day$^{-1}$). Then the profiles were summed for each trial
frequency offset.  The best fit frequency was determined by the Z$^2_6$ test 
(\cite{Buccheri83}) which measured the significance of the first 6 Fourier 
amplitudes. The frequency estimates were iteratively improved by generating a 
new phase model from the best fit frequencies and repeating the process. For 
GRO J2058+42,  root-mean-square (RMS) pulsed fluxes were estimated from the 
best fit pulse profile as 
\begin{equation}
F_{\rm RMS} = \left[ \int^1_0 (F(\phi) - \bar F )^2 d\phi \right]^{1/2}
\end{equation} 
where $F(\phi)$ is the pulse profile at phase $\phi$, $0 \leq \phi \leq 1$, and
$\bar F = \int^1_0 F(\phi)d\phi$ is the average flux.   The pulsed fluxes in 
the bottom panel of figure~\ref{fig:freqhis} were determined at 4-day intervals,
assuming an exponential energy spectrum, 
\begin{equation}
f(E) = A \exp(-\frac{E}{E_{\rm fold}})\label{eqn:spec}
\end{equation}
with a normalization $A$ and e-folding energy $E_{\rm fold}$ = 20 keV. 

For pulse periods longer than about 100 seconds, this frequency estimation 
method had an added complication if the pulse period was close to a harmonic 
of the spacecraft orbital period ($\approx$ 93 min). The pulse period of GRO 
J2058+42 was close the 28th spacecraft orbital harmonic. Relatively bright 
sources produced intensity steps in the count rate in source-facing detectors 
upon entering or exiting Earth occultation.  When the data from these detectors
were fit for a range of frequencies that included a spacecraft orbital harmonic,
the accumulated occultation steps produced a periodic signal which was 
sometimes more significant than the signal from GRO J2058+42. This effect was 
removed by discarding an interval of about 30 seconds of DISCLA data which was
centered on the occultation step of each interfering source and by allowing a 
corresponding discontinuity in the background model. Occultations for the Crab
and Cyg X-1 were always removed from the data. Also, occultations for GRS 
1915+105, Sco X-1, and GRO J1744-28 were removed when they were bright. 

From figure~\ref{fig:freqhis}, which shows the determined pulse frequency and 
pulsed flux history of GRO J2058+42 from BATSE data, it is apparent that GRO 
J2058+42 experienced a large spin up during its initial 46 day outburst. The 
period changed by about 2 seconds across this outburst, corresponding to an 
average $P/|\dot P| \approx$ 12 years. Measurements of the spin-up rate yielded
a peak value of $(2.48\pm .02)\times 10^{-11}$ Hz s$^{-1}$ (where we have
neglected corrections for the pulsar's orbital motion), which is comparable to
peak spin-up rates seen by BATSE in ``giant'' outbursts from known Be 
transients (\cite{Bildsten97}). In the bottom panel of figure~\ref{fig:freqhis}, 
the peak intensity of the initial outburst is about 10 times the peak intensity
of later outbursts.  The initial outburst begins at approximately the same
orbital phase as later outbursts, but peaks at a later phase. The large spin-up
and brightness of the initial outburst compared to later outbursts leads us to
classify it as a ``giant'' outburst (\cite{Stella86}). 

The total (pulsed+unpulsed) flux from GRO J2058+42 was determined using Earth
occultation measurements. This technique measures the intensity of a known 
source by calculating the difference in total count rate in source-facing 
detectors just before and just after occultation by the Earth 
(\cite{Harmon92}).  For sources with long pulse periods and large pulsed
fractions, flux measurements from individual occultation edges can be biased by
favoring a particular pulse phase.  However, this bias was easily eliminated by
averaging many occultation measurements to uniformly sample the pulse.

\begin{figure}
\psfig{file=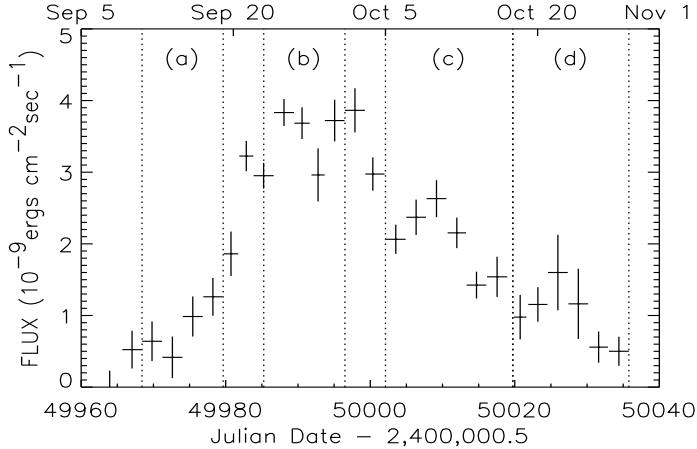,height=2.5in,width=4.0in}
\caption{GRO J2058+42 total 20-70 keV flux as measured by Earth
occultation.  The fluxes were determined at 2-day intervals by assuming an 
exponential spectrum with an e-folding energy of 20 keV. The weak outbursts were
not detected in Earth occultation data. The 3-$\sigma$ upper limit on the peak
fluxes for a 4-day interval was $\approx$ 50 mCrab. The labels (a),(b),(c), 
and (d) and dotted lines denote the intensity intervals used for the pulse 
profiles in figure~\protect\ref{fig:profile}.
\label{fig:occhis}}
\end{figure}

Both the Earth occultation measurements and the pulsed flux measurements were 
well fit by an exponential spectrum (see eq.~\ref{eqn:spec}) with $E_{\rm fold}$ 
values of 15-22 keV. No dependence of spectral shape on intensity was detected.
The total 20-70 keV flux history for GRO J2058+42 was determined at 2-day 
intervals by fitting the Earth occultation count rates with 
equation~(\ref{eqn:spec}) and assuming $E_{\rm fold} = 20$ keV.  This flux history
is shown in figure~\ref{fig:occhis}.  The weak outbursts were undetectable in 
Earth occultation data, with an upper limit of $\lesssim 50$ mCrab (3$\sigma$). 

Pulse profiles for 4 day intervals were generated by epoch-folding BATSE CONT
(2.048 s) data into 64 phase bins using a quadratic phase model (again
neglecting the pulsar's unknown orbit). Phase models were estimated at four day
intervals by a grid search in frequency and frequency derivative, as described
above. Each epoch-fold contained the usable data from a single spacecraft orbit.
Data were excluded when the GRO J2058+42 was occulted  and during intervals 
containing occultation steps for Crab and Cyg X-1. Gamma-ray bursts, Solar 
flares, South Atlantic Anomaly passages, and electron precipitation events in
the data stream were also excluded.

We searched for pulse shape variations using a statistical comparison of pulse 
profiles formulated in the frequency domain. Wilson et al. (1997) 
\nocite{Wilson97} show that the harmonic coefficients of a pulse profile are 
uncorrelated. For each single spacecraft orbit epoch-folded profile, these 
harmonic coefficients were calculated as 
\begin{equation}
a_k = \frac{1}{N}\sum^N_{j=1} R_j e^{-i 2 \pi jk/N} 
\end{equation}
where $k$ is the harmonic number, $ i = (-1)^{1/2}$, $N = 64$, 
the number of phase bins in the pulse profile, and $R_j$ is the count rate in 
the $j$th bin of the pulse profile.  
We calculated the mean value $\bar a_k$ from the measurements of $a_k$
from the epoch-folded profiles contained within each 4-day interval
corresponding to the phase model.  We calculated the error on the mean 
$\sigma_{\bar a_k}$ from the sample variance of the measurements of $a_k$. 
By comparing signal power to noise power $\bar a_k^2/\sigma_{\bar a_k}^2$, we determined
that only the first 6 harmonics were significant.  The average noise level was 
higher than expected for Poisson counting noise because Cyg X-1 was nearly always
present in the field of view and was in an active state during the giant
outburst of GRO J2058+42, introducing spurious noise power over a broad range
of frequencies. The observed noise power spectrum exhibited a turnover 
frequency consistent with Cyg X-1 (\cite{Crary96}).  Noise from Cyg X-1 also
affected the sensitivity of the frequency search.

\begin{figure}
\psfig{file=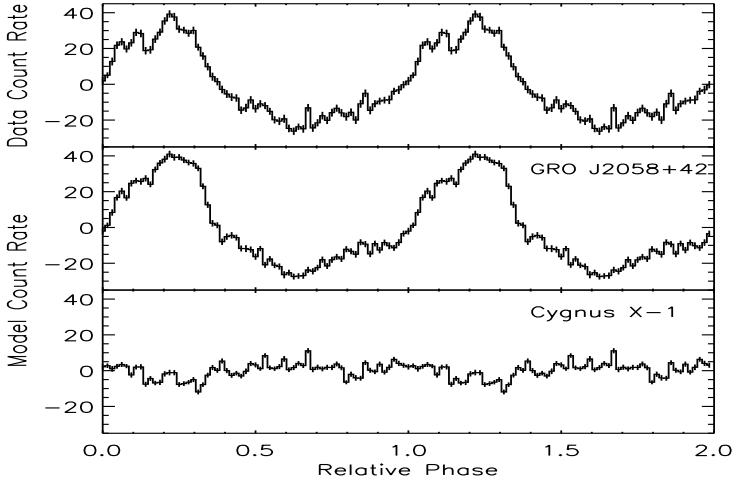,height=2.5in,width=4.0in}
\caption{Noise effects in the GRO J2058+42 pulse profile produced by Cygnus
X-1. The top panel is the epoch-folded 20-70 keV pulse profile for 1995 
September 23-27. The center panel is the modeled count rate for GRO J2058+42, 
assuming an exponential spectrum (see eq.~\protect\ref{eqn:spec}) with an e-folding
energy of 20 keV. The bottom panel is the modeled count rate for Cygnus X-1 
with a power law spectrum of photon index of -1.85.  The count rate from Cygnus
X-1 averaged near zero because we measured only the effects of noise from 
Cygnus X-1 on the mean subtracted profile.
\label{fig:cyg}}
\end{figure}

We attempted to remove Cyg X-1 in the time domain by simultaneously fitting 
the two sources using different assumed spectra. For each phase bin, fluxes 
were determined for GRO J2058+42 and Cyg X-1 by minimizing 
\begin{equation}
\chi^2 = \sum_{i=1}^6 \frac{(\Delta C_i - \alpha_i S - \beta_i N)^2}
{\sigma_i^2}
\end{equation}
where $\Delta C_i$ is the mean subtracted count rate for the $i$th energy channel
in a 4-day mean profile, $S$ is the GRO J2058+42 flux at 30 keV, and $N$ is the
Cyg X-1 flux at 30 keV. The parameter $\alpha_j$ is the assumed energy spectrum
for GRO J2058+42, $f(E) = \exp(-(E-30$ keV)$/20$ keV$)$, convolved with the 
detector response to yield the expected count rate in energy channel $i$. 
The parameter $\beta_i$ was calculated in the same way using a photon power law
spectrum with photon index -1.85 for Cyg X-1.  The energy channel
range used corresponded to 20-125 keV. GRO 2058+42 was detectable up to about 
70 keV while Cyg X-1 emission was present up to about 400 keV.  This method 
worked well for only the brightest profiles, due to statistical limitations.
Figure~\ref{fig:cyg} shows the fit to a profile from 1995 September 23-27
near the peak of the giant outburst. Noise from Cyg X-1 appears to be causing
high frequency features in the profile. 

Since noise from Cyg X-1 could not be directly removed, only the first 6 
harmonics of the pulse profiles were retained for pulse shape analysis, thus
avoiding most of the noise effects.  Groups of harmonic limited 4-day profiles
were phase-aligned and averaged to improve statistical significance. Mean 
profiles from different epochs were compared by minimizing        
\begin{equation}
\chi^2 = \sum_{k=1}^6 \frac{|\bar p_k - \bar t_k \beta e^{2 \pi i \Delta \phi}|^2}
 {\sigma_{\bar p_k}^2 + \beta^2 \sigma_{\bar t_k}^2}
\end{equation}
where $\bar p_k$ and $\bar t_k$ were the harmonic coefficients of the first and
second mean profiles respectively and $\sigma_{\bar p_k}$ and $\sigma_{\bar
t_k}$ were the corresponding errors. The second mean profile was scaled by a 
factor $\beta$ and shifted in phase by $\Delta \phi$ to obtain the best fit.  

\begin{figure}
\psfig{file=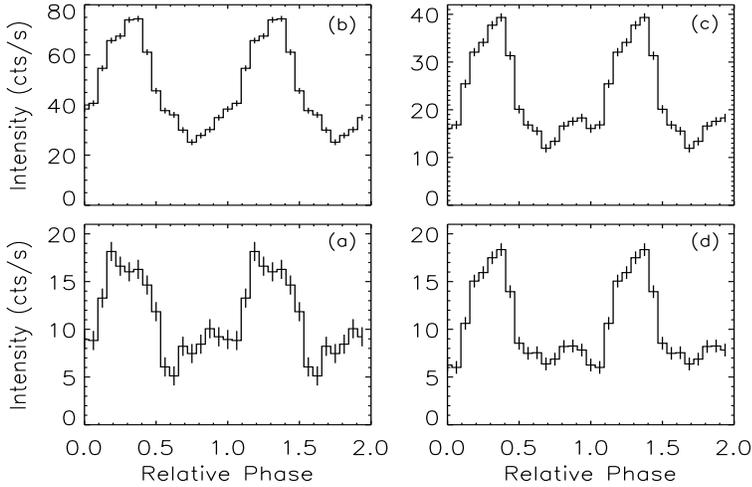,height=2.5in,width=4.0in}
\caption{Phase aligned pulse profiles from 20-70 keV BATSE CONT data for the
4 intervals, (a) 1995 
September 11-18, (b) September 23-30, (c) October 5-17, and (d) October 17-28.
The average count rate from BATSE Earth occultation data from corresponding 
time intervals (figure~\protect\ref{fig:occhis}) was added to the count rate 
plotted to demonstrate the pulsed fraction and relative intensity of the 
profiles, but was not been in the errors. The outburst proceeds clockwise from
lower left.
\label{fig:profile}}
\end{figure}

Phase aligned pulse profiles were thus constructed from BATSE 20-70 keV CONT
data from four intervals: (a)1995 September
11-18,(b) September 23-30, (c) October 5-17, and (d) October 17-28.
The intensity intervals used to construct these profiles are denoted by dotted 
lines in figure~\ref{fig:occhis} and are labeled accordingly. The profiles
themselves are shown in figure~\ref{fig:profile}. Average count rates measured
by Earth occultation for these time intervals were added to the profiles shown
to demonstrate the relative intensities of the profiles. Errors on these count
rates were not included in the errors plotted because the profile comparisons 
involved only the mean subtracted profiles.  The peak-to-peak pulse fraction,
$(\bar r_{\rm pulsed} - r_{\rm pulsed_{\rm min}})/r_{\rm occ}$, for these profiles was 
consistent with a constant value of 0.46$\pm$ 0.02.  

The profiles near the peak  of the outburst (figure~\ref{fig:profile}b) were 
significantly different from profiles at lower intensities ($\lesssim$ 100 
mCrab). The main pulse appeared to evolve from a faster rise than fall to a 
slower rise than fall, although the primary differences in the profiles were 
occurring in the region between the large pulses (phase 0.5-1).  Profile (a) was
relatively flat in the interpulse region. The deep dip near phase 0.6 persisted
in higher channels and was most likely due to Cyg X-1.  Profile (b) slowly 
declined in intensity to a minimum near phase 0.7, after which it slowly 
increased in intensity. Profiles (c) and (d) both had a feature from phase 
0.8-1.0 that did not persist at higher energies, so it was most likely 
intrinsic to the source.  The profile in (a) was significantly different from 
the profile in (d), although the intensities were similar. However, profiles 
(c) and (d), both from the fall of the outburst were consistent with one 
another, although their intensities differed. Profiles during the rise of the
outburst differed from those in the fall of the outburst which implies that 
the shape of the profile was dependent upon more than simply the mass 
accretion rate.   

\begin{figure}
\psfig{file=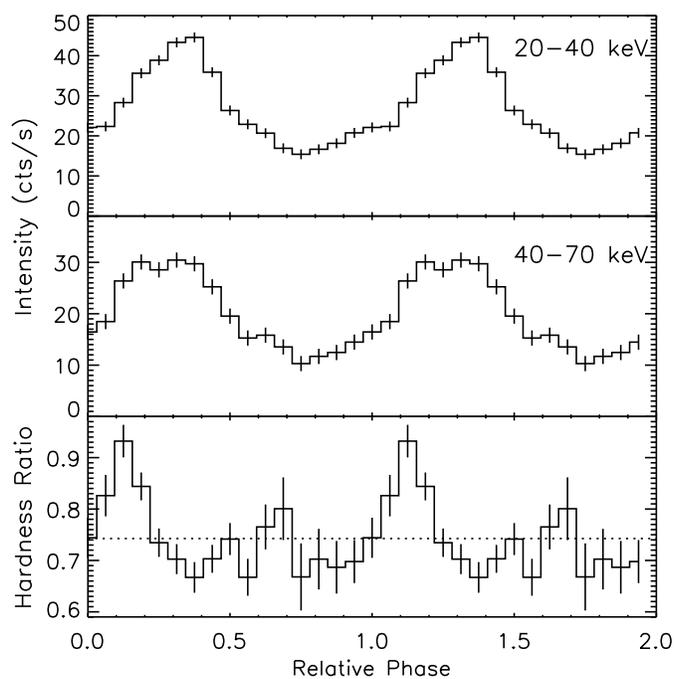,height=3.5in,width=3.5in}
\caption{Energy dependent pulse profile variations for 1995 September
23-30.  The top and center panels are the pulse profiles from 20-40 keV
and 40-70 keV generated by epoch-folding BATSE CONT data. Average count rates
and errors from the corresponding Earth occultation data were added to the 
epoch-folded profiles.  The bottom panel is the ratio of the intensity in the
center panel to the top panel. Errors are for the difference between this 
ratio and the average hardness ratio. The dotted line corresponds to the 
average hardness ratio, 0.74$\pm$0.06, calculated from Earth occultation data 
only.
\label{fig:hardness}}
\end{figure}

Next we compared profiles for the energy intervals 20-40 keV and 40-70 keV.  We 
detected profile variations with energy as shown in figure~\ref{fig:hardness}.
The profiles shown are from the brightest interval 1995 September 23-30.
Comparisons from fainter intervals were also suggestive of energy dependent 
pulse shape variations. In figure~\ref{fig:hardness}, average count rates and
errors from Earth occultation measurements were added to the profiles. The 
bottom panel is the hardness ratio given by 
\begin{equation}
r = \frac{h+\Delta h}{s+\Delta s}
\end{equation}
where $h$ is the average count rate measured by Earth occultation for 40-70
keV, $\Delta h$ is the mean subtracted pulse profile count rate for 40-70 keV,
$s$ is the average count rate for 20-40 keV, and $\Delta s$ is the mean
subtracted profile for 20-40 keV. The errors shown are the errors on the
difference $r - h/s$, since we are interested in where the hardness ratio
deviated from the mean value, 0.74 $\pm$ 0.06. The hardness ratio demonstrated
that the rise of the pulse (phase 0.0-0.25) had a harder spectrum than later
pulse phases.  

\subsection{Location with {\em RXTE}}

\begin{figure}
\psfig{file=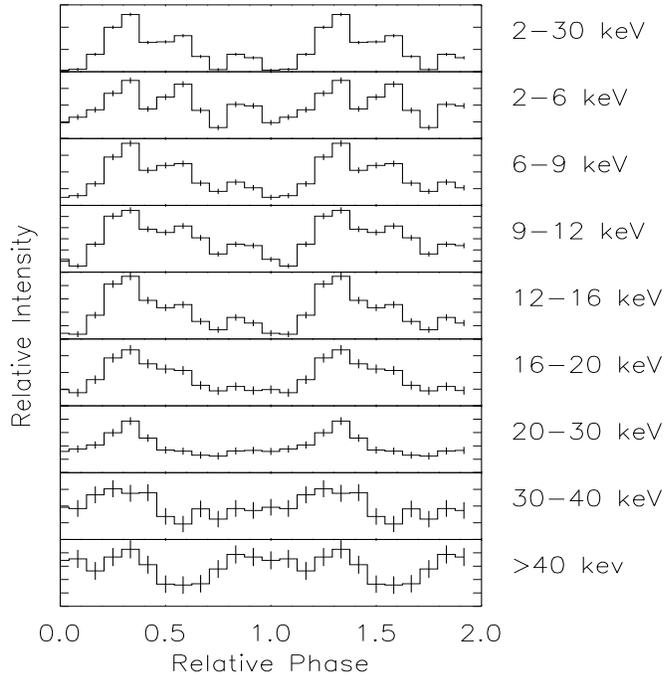,height=3.5in,width=3.15in}
\caption{Pulse profiles from the pointed portion of the {\em RXTE} PCA
observation on 1996 November 28. The pulse shape appears to be changing with 
energy.
\label{fig:xte_profiles}}
\end{figure}

The Proportional Counter Array (PCA) on {\em RXTE} consists of five proportional counters sensitive to
photons from 2-60 keV and has a total collecting area of 6500 cm$^2$. The PCA
is a collimated instrument with an approximately circular collimator efficiency
with a FWHM of about 1\arcdeg\ (\cite{Jahoda96}). 
PCA observations of GRO J2058+42 were performed on 1996 November 28. A
series of scanning observations was followed by a short ($\approx 900 s$)
dwell on the source. A frequency consistent with the BATSE barycentered
measurement,  $\nu = 5.1153 \pm 0.0003 \times 10^{-3}$ Hz at MJD 10416, and
three higher harmonics were clearly detected during the pointed observation.
Figure~\ref{fig:xte_profiles} shows the pulse profiles as a function of
energy from the PCA data.  The pulse shape appears to be changing with energy.
Evidence of variations can be seen by comparing the feature near phase 0.6 in
different energy ranges.  This feature is quite prominent at lower energies but
is essentially nonexistent above 20 keV.  

\begin{figure}
\psfig{file=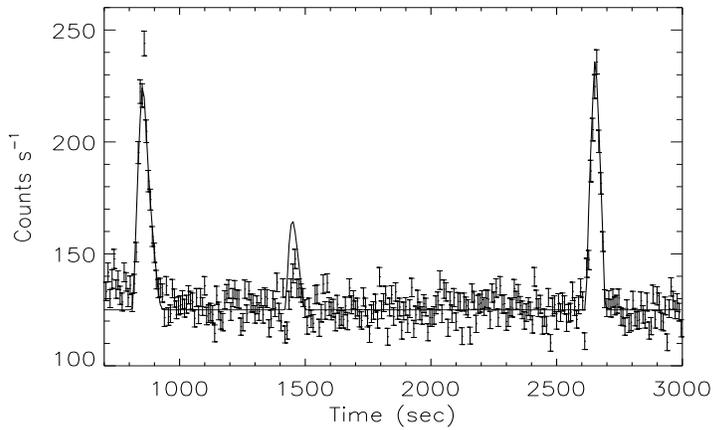,height=2.5in,width=4.0in}
\caption{The observed count rate and best fit model for the scan portion of
the {\em RXTE} PCA observation used to derive the position of GRO J2058+42. The
peaks result from scans over and/or near the position of GRO J2058+42.
\label{fig:pca_mod} }      
\end{figure}

To estimate the position of
GRO J2058+42 from the PCA scanning data, we analyzed the two largest scan peaks
simultaneously. The relatively constant background during this
time greatly simplified the analysis. We used the spacecraft pointing data
combined with a model of the collimator response to calculate the contribution
to the total 2-60 keV count rate from a constant (in time) background $B$ and a
model source of assumed constant intensity $S$ (counts s$^{-1}$) located at
celestial coordinates $\alpha$, and $\delta$. We minimized the four parameter
function 
\begin{equation}
\chi^2 = \sum_i \frac{( O_i - M_i(S,\alpha_j,\delta_j , B))^2}{\sigma^2_i}. 
\end{equation}
Here $O_i$, $M_i$ are the observed and predicted count rates for time interval $i$, and $\sigma^2_i$ is the variance on the
predicted count rate. The best-fit source position was that which had the minimum
value for $\chi^2$. We determined a statistical confidence region from the 
appropriate value of $\Delta\chi^2$ for the 3 interesting parameters ($S,
\alpha_j, \delta_j$.) To the extent
that the background was flat the background parameter was not relevant to the 
determination of the confidence region (see \cite{Lampton76}). 
We show in figure~\ref{fig:pca_mod} the data and best-fit model. The resulting
best-fit position was $\alpha = 20^h 59^m$, $\delta$ = 41\arcdeg 43\arcmin\ 
(\cite{Wilson96}). The total uncertainty in the position was not
dominated by the statistical error, rather, time variability of the source as
well as uncertainties in the collimator model contributed to the error budget as
well. Our current understanding of these effects yielded a final circular
confidence region (90 \%) of radius 4\arcmin\ centered at the above position.

\subsection{The Weak Outbursts}

\begin{figure}
\psfig{file=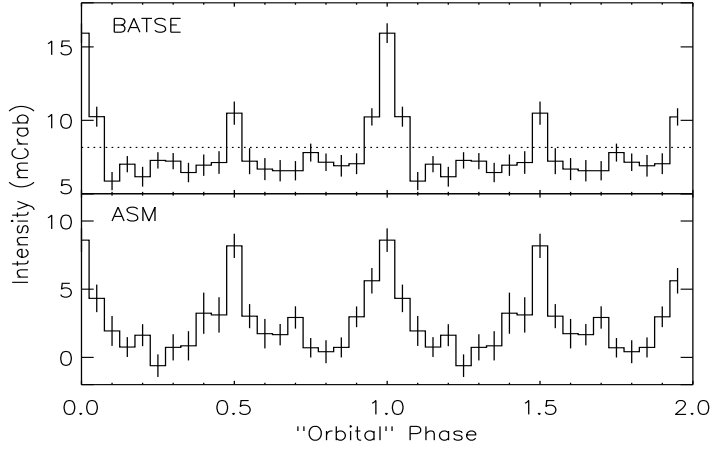,height=2.5in,width=4.0in}
\caption{The BATSE (20-50 keV) pulsed flux and
the ASM flux from the first 7 weak outbursts were epoch-folded at a period of 110 
days (epoch Julian Date 2450302).  The eighth outburst was excluded because noise
from Cyg X-1 affected the BATSE measurements.  The average 99\% confidence
upper limit on the BATSE pulsed flux is denoted by a dotted line.
\label{fig:orb_profile}}
\end{figure}

The {\em RXTE} All-Sky Monitor (ASM) consists of three wide-angle shadow 
cameras equipped with Xenon proportional counters with a total collecting area
of 90 cm$^2$.  The ASM provides 90 second snapshots of most of the sky every 
96 minutes in three energy channels from 2-12 keV (\cite{Levine96}). Archival 
and new image data at the position of GRO J2058+42 determined by the PCA were 
searched by the {\em RXTE} ASM team and weak outbursts were detected. Single
dwell (90 s time resolution) data during these outbursts were epoch-folded
using phase models determined from BATSE data, but no pulsations were detected.
Figure~\ref{fig:xte} is a comparison of the 
BATSE pulsed flux (20-50 keV) and the flux as measured by the {\em RXTE} ASM.
The BATSE data indicated significant detections every $\approx$110 days of outbursts 
peaking at 15-20 mCrab (20-50 keV). Halfway between these outbursts BATSE 
detected shorter weaker outbursts with peak fluxes $<$ 15 mCrab. The ASM data 
show outbursts every 54 days (\cite{Corbet97}) that were all similar in 
intensity and duration. Figure~\ref{fig:orb_profile} shows BATSE pulsed flux 
and ASM flux from the first 7 weak outbursts folded at a period of 110 days 
with epoch Julian Date 2450302. The folding epoch was chosen near the peak of a
stronger outburst in the BATSE data. The outbursts near phase 0.0 were much 
brighter in the BATSE data than those near phase 0.5, while both outbursts were
similar in the ASM data.  No known systematic effects would remove or weaken 
outbursts in the BATSE data at 55 day intervals. A closer analysis of the
individual {\em RXTE} ASM energy channels did not show a significant difference
in the spectral hardness of the two sets of outbursts within the 2-12 keV band.

\section{Discussion}

The optically identified companions for transient pulsars with pulse periods 
longer than one second are all Be or Oe stars.  Long term studies of these Be/X-ray 
binaries have demonstrated that giant outbursts followed by or interspersed with
a series of periodic normal outbursts appear to be typical behavior for 
Be/X-ray binaries (\cite{Bildsten97}).
GRO J2058+42 is most likely a Be/X-ray pulsar because it exhibits transient 
outbursts recurring with a presumed orbital period of 110 days and it has both
giant and normal outbursts. An orbital period of 110 days places GRO J2058+42
along the orbital period spin period correlation for Be/X-ray binaries 
(\cite{Corbet86}, \cite{Waters89a}.) 

The giant
outburst of GRO J2058+42 showed enough dynamic range that we can use the
relationship between torque and observed flux to test accretion theory.
Simple accretion theory assumes that material from the companion star is
flowing onto a rotating neutron star with a magnetic field.
The magnetic field is so strong that it determines the motion of the material 
in a region of space surrounding the neutron star called the magnetosphere. 
The size of this region, the magnetospheric radius $r_{\rm m}$, is defined to be the
distance from the neutron star at which all magnetic field lines are just 
closed loops.  Another important length scale is the corotation radius, 
$r_{\rm co}$, the distance from the neutron star where centrifugal forces just 
balance local gravity.  If $r_{\rm m} > r_{\rm co}$, accretion is not expected to occur.
Accretion theory predicts $r_{\rm m} \propto \dot M^{-2/7}$ for $r_{\rm m} < r_{\rm co}$ for
for disk or wind accretion. The accretion torque $N$ is given
by $N \propto \dot M \sqrt r_{\rm m}$.  We can assume the bolometric flux
$F_{\rm bol}$
is related to the mass accretion rate $\dot M$ by $\dot M \propto F_{\rm bol}$.
Therefore, simple accretion theory predicts $\dot \nu \propto F_{\rm bol}^{6/7}$,
where $\dot \nu$ is the spin-up rate of the neutron star (\cite{Henrichs83} and
references therein). 

\begin{figure}
\psfig{file=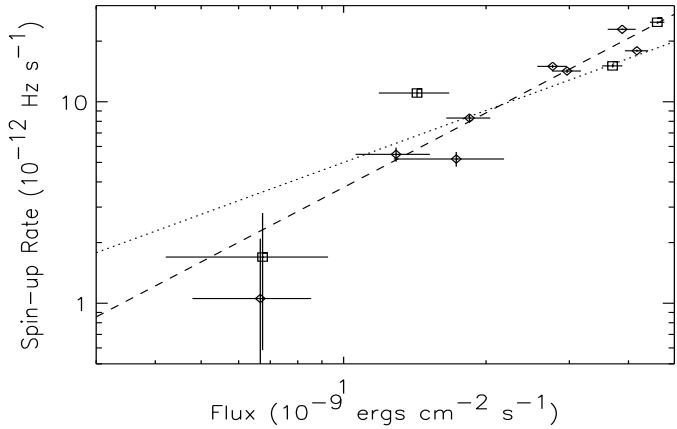,height=2.5in,width=4.0in}
\caption{The observed 20-100 keV flux as measured by BATSE Earth occultation
is plotted versus the measured pulsar spin-up rate $\dot \nu$ during the giant
outburst. The square symbols indicate the rise of the outburst and the diamond
symbols denote the decline. The dotted curve is a power law with an index of 
6/7, the relationship between total flux and $\dot \nu$ predicted by accretion
torque theory. The dashed curve is the best fit power law with an index of 1.2.  
The weak outbursts place an upper limit of 5$\times 10^{-12}$ Hz s$^{-1}$ on the orbital contribution
to $\dot \nu$.\label{fig:flvsfd}}
\end{figure}

Figure~\ref{fig:flvsfd} shows the observed 20-100 keV flux
as measured by Earth occultation plotted versus the measured frequency 
derivative $\dot \nu$ during the giant outburst. The frequency derivatives were
generated by an search performed on the BATSE DISCLA 20-50 keV 
data over a grid of trial frequencies and frequency derivatives. This technique
was identical to that described in section~\ref{sec:giant} except
that the pulse profiles were shifted according to a grid of frequency offsets and 
frequency derivatives, rather than only frequency offsets. The search 
yielded maximum spin-up rates of 5 $\times 10^{-12}$ Hz s$^{-1}$ in the weak 
outbursts.  If we assume that the orbital contribution was small compared to the
intrinsic $\dot \nu$, we can treat the measured $\dot \nu$ as equal to the 
intrinsic $\dot \nu$ and compare it to the total flux in the BATSE energy range.
Clearly, $\dot \nu$ and the total flux were correlated, hence an accretion disk
was likely to be present. The power law $\dot \nu \propto F^{\gamma}_{\rm obs}$ with
$\gamma = 6/7$, which is predicted by accretion torque theory, is shown in 
figure~\ref{fig:flvsfd}. Also shown is the best fit curve with $\gamma \simeq 
1.2$. Orbital contributions to the torque and the fact that BATSE
does not measure the bolometric flux may explain the discrepancy between the 
measured slope and accretion theory. Interestingly, a similar fit to EXOSAT data
from the 1985 giant outburst of the Be/X-ray binary EXO 2030+375 also gave $\gamma \simeq 
1.2$ (\cite{Reynolds96}).

In addition to testing accretion theory, we can use the maximum 
$\dot \nu$ ($\dot \nu_{\rm max} = 2.48 \times 10^{-11}$ Hz s$^{-1}$) from
the giant outburst to estimate a distance to GRO J2058+42.  The luminosity is
related to $\dot \nu$ by 
\begin{equation}
L_{37} = 4.83 \times 10^{13} \mu_{30}^{-1/3} m^{1/2} R_6^{-1} I_{45}^{7/6}
n(\omega_s)^{-7/6} \dot \nu^{7/6} 
\end{equation} 
where the notation is $\mu_{30} = \mu/10^{30}$ G cm$^3$, $ m =
M_x/M_{\sun}$, $R_6 = R/10^6$ cm, $I_{45} = I/10^{45}$ g cm$^2$, and
$L_{37} = L_{\rm acc}/10^{37}$ ergs s$^{-1}$.  The neutron star has magnetic moment
$\mu$, radius R, mass M$_x$, moment of inertia I, accretion luminosity
L$_{\rm acc}$, and ``dimensionless torque" $n(\omega_s)$.  We assumed the very slow
rotator case of $n(\omega_s) \approx 1.4$ (\cite{Henrichs83} and references
therein).  We used typical values (m = 1.4, $R_6 = 1$,
$I_{45} = 1$). Since $\mu_{30}$ is less well known, we calculated bolometric 
luminosities, L$_{\rm acc} \simeq 0.7-3.5 \times 10^{38}$ ergs s$^{-1}$,
for the range $\mu_{30} = 0.1 - 10$. These luminosities were comparable to the 
Eddington limit for neutron stars, L$_{\rm Edd} \simeq 2 \times 10^{38}$ ergs
s$^{-1}$. Next, to calculate the distance, we can 
use the 20-100 keV flux corresponding to $\dot \nu_{\rm max}$, 
$F_{\rm max} = 4.76\times10^{-9}$ ergs cm$^{-2}$ s$^{-1}$. Lastly, we must
calculate a bolometric correction $\alpha$, the ratio between the BATSE flux 
and the total 2-100 keV flux.  Simultaneous BATSE and {\em RXTE} measurements from 1996
November yielded a bolometric correction of $\alpha \simeq 0.4$. Then the 
distance $d$ is given by 
\begin{equation}
 d = \frac{1}{3.09\times10^{21} \mathrm{cm}} (\frac{\alpha L_{\rm acc}}{4\pi 
F_{\rm max}})^{1/2} \mathrm{kpc}
\end{equation}
which gives distances of 7-16 kpc for GRO J2058+42. For these distances the
estimated bolometric luminosity for the 1996 November outburst is L$_{\rm bol}
\simeq 0.2-1 \times 10^{37}$ ergs s$^{-1}$. Bolometric luminosities for the 
intermediate weaker outbursts (in the BATSE data) appear to be similar to the
1996 November outburst, but are more uncertain because complete spectral
and pulsed fraction information are unavailable.
 
\begin{figure}
\psfig{file=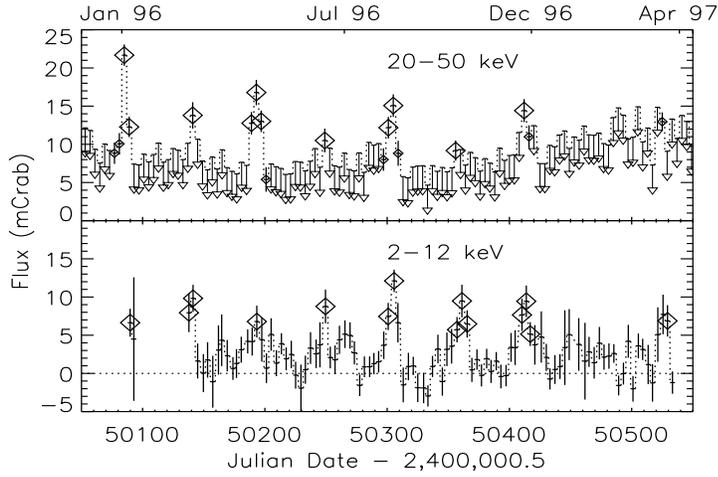,height=2.5in,width=4.0in}
\caption{The top panel is BATSE (20-50 keV) pulsed flux where 1 mCrab = 6.3
$\times 10^{-3}$ keV cm$^{-2}$ s$^{-1}$.  The large diamonds denote detections at the
99.9\% confidence level.  The small diamonds denote less significant
measurements that are considered likely detections because the measured
frequencies are well aligned with those that meet the 99.9\% criterion.  The
likely detection for 1997 March 16-20 was confirmed by a more sensitive
search in frequency and frequency derivative.  Arrows denote 99\% upper limits
for the pulsed flux.  The reduction in sensitivity from 1996 December is due to
increased noise from Cygnus X-1. The bottom panel is the total flux as measured
by the {\em RXTE} ASM (2-12 keV) where 1 mCrab = 0.075 counts s$^{-1}$. The
{\em RXTE} ASM flux
has been binned into 4-day intervals. Points in the {\em RXTE} dataset where the 
count rate is at least 3 times the error are marked with large diamonds.
\label{fig:xte}}  
\end{figure}

The weak outbursts provide interesting, but difficult to interpret, information
about this system. The BATSE data indicate brighter and longer outbursts every
$\approx$110 days with weaker outbursts in between, while the ASM data show outbursts 
every 54 days that are similar in intensity and duration (fig~\ref{fig:xte} \& 
\cite{Corbet97}). The intermediate weaker outbursts in the BATSE data must be 
of a different character than the stronger outbursts.  The fact that BATSE 
detects a 110 day periodicity and {\em RXTE} detects a periodicity at half that
period implies that either a spectral change or a change in the pulse fraction
is occurring every other outburst, indicating a difference in accretion mode.
An analysis of the three {\em RXTE} ASM energy bands shows that both sets of 
outbursts have comparable intensity in the 5 - 12 keV band. This implies that 
if a spectral change is occurring it is above 12 keV.
The observed periodicity may be interpreted in two ways: (1) the 110 day cycle 
observed in the BATSE data is the orbital period and 2 outbursts are occurring 
each orbit; or (2) the orbital period is 54 days and the spectrum (or pulse
fraction) alternates every other orbit for at least 9 orbits. We can find no
plausible explanation for an alternating change in spectrum or pulse fraction
that persists for at least 9 orbits if the outbursts are produced by periastron
passage in a 54 day orbit. Therefore we interpret the observed 110 day 
periodicity as the orbital period. 

We propose that GRO J2058+42 
is undergoing periastron and apastron outbursts in a 110 day orbit. 
A neutron star in an inclined orbit combined with concentration of material in
the equatorial plane of the Be star is a possible explanation for the outburst
behavior observed in GRO J2058+42.  The same velocity kicks that produce an 
eccentric orbit could also produce an inclined orbit. Outbursts would occur 
twice per orbit when the neutron star crosses the equatorial plane of the 
companion (\cite{Priedhorsky87}).  From the BATSE data, we measured the separation between the 
brighter outbursts to be 110 $\pm$ 3 days  and the separation between brighter
and weaker outbursts to be $\Delta T = 54.5 \pm$ 2.3 days.  If we assume the 
separation between the brighter outbursts is the orbital period, $P_{\rm orb}$, then
$\Delta T/P_{\rm orb}$ falls in the 95\% confidence interval 
$0.45 < \Delta T/P_{\rm orb} < 0.55$.  For this interval the separation
between periastron and the line where the neutron star's orbit intersects the 
equatorial plane of the companion is $\nu <$ 12.1\arcdeg\ for a typical 
eccentricity, $e = 0.35$.  Hence this mechanism would produce outbursts near periastron and near
apastron. This mechanism does not, however, explain the intensity differences 
seen only in the higher energies (20-50 keV).  Although an inclined orbit has 
often been proposed as a mechanism for outbursts in Be/X-ray binaries, two
outbursts per orbit have not been seen in other Be/X-ray binaries to date.
However, such behavior has been observed in wind-fed systems such as GX 301-2
(\cite{Koh97}).
 
Another explanation which does not require an inclined orbit suggests that two
different accretion mechanisms could be at work in this system. An accretion
disk is likely to be present during giant outbursts. Its presence helps to
explain the large and steady spin-up rates seen during giant outbursts. A peak
spin-up rate of 2.5 $\times 10^{-11}$ Hz s$^{-1}$ was observed in GRO J2058+42.
Bildsten et al. (1997) suggest that an accretion disk could remain after a 
giant outburst.  Then normal outbursts could be produced by large tidal torques
on the accretion disk during periastron passage.  This could explain the series
of normal outbursts following the giant outburst. The apastron outbursts could
be produced by accretion from the equatorial outflow from the Be star.  A slow
dense outflow could produce outbursts at apastron (\cite{Waters89b}).  The two outbursts
per orbit would be produced by different accretion mechanisms which could 
produce differences in spectra or in pulse fraction. 
Clearly, additional observations are needed to fully understand this source.

\acknowledgements{ The quick-look ASM data on GRO J2058+42 were provided by the
{\em RXTE}/ASM team at MIT and NASA/Goddard Space Flight Center. We thank Jan 
van Paradijs and Matthew Scott for helpful discussions. D.~C. was a NASA 
Compton Postdoctoral Fellowship under grant NAG 5-3109. We thank the referee 
L.~B.~F. (Rens) Waters for helpful comments.}

\end{document}